\documentclass[conference]{IEEEtran}
\usepackage[pdftex]{graphicx}
\usepackage{comment}

\ifCLASSINFOpdf
\else
\fi
\usepackage[font=footnotesize]{subfig}
\usepackage{color}
\usepackage{framed}
\usepackage{comment}
\usepackage{hyperref}

\newcounter{t0d0_counter}
\newcommand{\nofixme}[1]{
}
\newcommand{\fixme}[1]{
 \stepcounter{t0d0_counter}
 \definecolor{shadecolor}{rgb}{1,1,0} 
 \begin{shaded}
 T0D0 \arabic{t0d0_counter}: #1
 \end{shaded}
}

\usepackage[switch]{lineno}

\newcommand{\dumpFourZeroSix}{\texttt{dump406}}
\newcommand{\dumpTenNinety}{\texttt{dump1090}}

\begin{document}


%
\title{Cybersecurity of COSPAS-SARSAT and EPIRB: threat and attacker models, exploits, future research}


\author{
\IEEEauthorblockN{Andrei Costin, Syed Khandker, Hannu Turtiainen, Timo H\"{a}m\"{a}l\"{a}inen}
\IEEEauthorblockA{Faculty of Information Technology \\
University of Jyv\"{a}skyl\"{a} \\
Finland \\ 
\{ancostin,syibkhan,turthzu,timoh\}@jyu.fi}
}



%


\IEEEoverridecommandlockouts
\makeatletter\def\@IEEEpubidpullup{6.5\baselineskip}\makeatother
\IEEEpubid{\parbox{\columnwidth}{
    \textit{AUTHORS' PREPRINT, DRAFT ACCEPTED BY REVIEWERS} of Workshop on Security of Space and Satellite Systems (SpaceSec)\\
    Network and Distributed System Security (NDSS) Symposium 2023\\
    28 February - 4 March 2023, San Diego, CA, USA\\
    ISBN 1-891562-83-5\\
    https://dx.doi.org/10.14722/ndss.2023.23xxx\\
    www.ndss-symposium.org
}
\hspace{\columnsep}\makebox[\columnwidth]{}}

\maketitle

\begin{abstract}

COSPAS-SARSAT is an International programme for ``Search and Rescue'' (SAR) missions based on the ``Satellite Aided Tracking'' system (SARSAT).
It is designed to provide accurate, timely, and reliable distress alert and location data to help SAR authorities of participating countries to assist persons and vessels in distress. Two types of satellite constellations serve COSPAS-SARSAT, low earth orbit search and rescue (LEOSAR) and geostationary orbiting search and rescue (GEOSAR). 
Despite its nearly-global deployment and critical importance, unfortunately enough, we found that COSPAS-SARSAT protocols and standard 406 MHz transmissions lack essential means of cybersecurity. 

In this paper, we investigate the cybersecurity aspects of COSPAS-SARSAT space-/satellite-based systems. 
In particular, we practically and successfully implement and demonstrate the first (to our knowledge) attacks on COSPAS-SARSAT 406 MHz protocols, namely replay, spoofing, and protocol fuzzing on EPIRB protocols. 
We also identify a set of core research challenges preventing more effective cybersecurity research in the field and outline the main cybersecurity weaknesses and possible mitigations to increase the system's cybersecurity level. 

\end{abstract}



%

\section{Introduction}

COSPAS-SARSAT is an International programme for ``Search and Rescue'' (SAR) missions that are based on ``Satellite Aided Tracking'' system (SARSAT)~\cite{ahmed2007satellite}.
It is organized as a treaty-based, nonprofit, intergovernmental, humanitarian cooperative of 45 nations and agencies~\cite{cospas-participants,Levesque93COSPAS}. 
COSPAS stands for ``Cosmicheskaya Sistema Poiska Avariynyh Sudov''
, which translates from russian as ``Space system for the search of vessels in distress''.
It is designed to provides accurate, timely, and reliable distress alert and location data to help SAR authorities of participating countries to assist persons, vessels and aircraft in distress. 

Despite being a long-running and highly-critical system (both from space/satellites and SAR points of view), to the best of our knowledge at present there are no public nor peer-reviewed works that investigate threat/attacker models or demonstrate practical attacks on COSPAS-SARSAT in general, and EPIRB in particular. 
In this paper we try to close several gaps, therefore our contributions are as follows:

\begin{enumerate}

\item We are the first (to the best of our knowledge) to approach and research the cybersecurity aspects of COSPAS-SARSAT systems, namely threat and attacker models, and future research directions.

\item We implement and present the first (to the best of our knowledge) practical attacks (e.g., spoofing) on COSPAS-SARSAT, and specifically EPIRB implementations more specifically.

\item We develop and plan to release \dumpFourZeroSix{} -- possibly the first implementation of open-source receiving and decoding software for COSPAS-SARSAT EPIRB 406 MHz beacon distress, an equivalent of the famous \dumpTenNinety{}~\cite{sanfilippo2014dump1090} widely used in ADS-B crowdsourcing and research communities.

\end{enumerate}

The rest of this paper is organized as follows.
We introduce basic background knowledge on COSPAS and SARSAT in Section~\ref{sec:background}. 
In Section~\ref{sec:relwork}, we discuss related work and state-of-the-art. 
Then we present the methodology and implementation details in Section~\ref{sec:meth-impl}. 
Following this, we detail the models for the attacker, threats, and exploits in Section~\ref{sec:attacks}. 
We then discuss the core challenges and future work in Section~\ref{sec:discuss}. 
Finally, we conclude the paper with Section~\ref{sec:concl}.


\section{Background}
\label{sec:background}

Two types of satellite constellations serve COSPAS-SARSAT, low earth orbit search and rescue (LEOSAR) and geostationary orbiting search and rescue (GEOSAR). The LEOSAR satellite constellation has five satellites, having an approximate orbital period of 100 minutes. When the LEOSAR system detects a distress alert, it calculates the location of the distress event using Doppler processing techniques and then forwards that data later when it passes into view of a ground station. Four GEOSAR satellites remain stationary in orbit relative to the Earth. Upon receiving any beacon signal, they relay the distress message. 
COSPAS-SARSAT supports three different types of beacon systems, namely Emergency Locator Transmitter (ELT)~\cite{ELT}, Personal Locator Beacon (PLB)~\cite{PLB}, and Emergency Position-Indicating Radio Beacon (EPIRB)~\cite{EPIRB}. 

\begin{figure*}[!htb]
\centering
\includegraphics[width=0.80\textwidth]{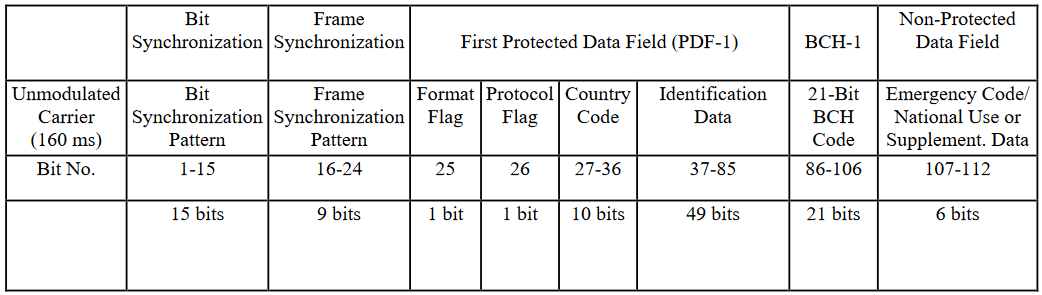}
\caption{Structure of the 406 MHz beacon \emph{short message} according to~\cite{cospas_specification}.}
\label{fig:short_message_structure}
\end{figure*}

\begin{figure*}[!htb]
\centering
\includegraphics[width=0.80\textwidth]{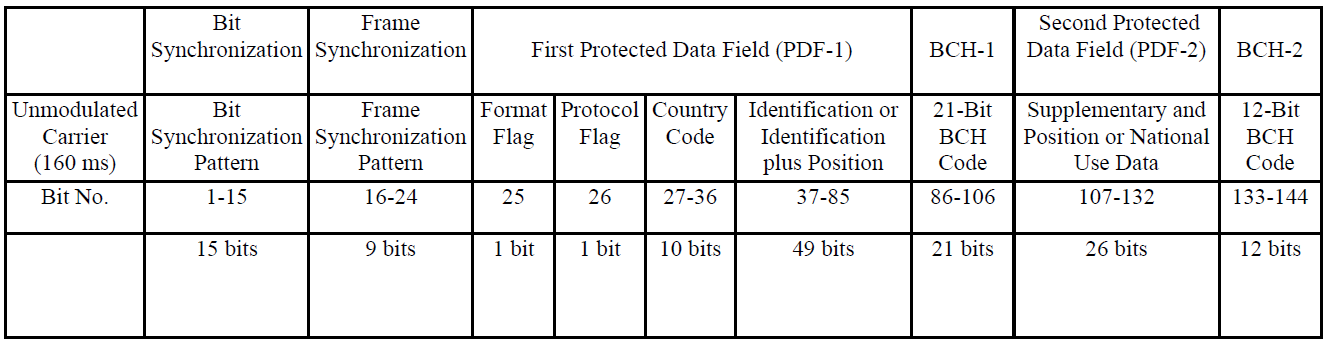}
\caption{Structure of the 406 MHz beacon \emph{long message} according to~\cite{cospas_specification}.}
\label{fig:long_message_structure}
\end{figure*}

The ELTs are mainly used by aircraft. This device was designed to be activated automatically or manually when a plane experiences physical shock (e.g., crash) or comes in the touch of water. Early ELTs used analog signals on 121.5 MHz or 243 MHz. However, since February 2009, COSPAS-SARSAT has supported ELT reception only at 406 MHz to improve the service quality and synchronize with other beacon systems. An ELT signal consists of a 160 ms unmodulated carrier followed by a 280 ms (short message 112 bits) or 360 ms (long message 144 bits) digitally modulated carrier signal. From 25-bit to 85-bit (61 bits segment) is called a protected data field that contains the primary data (e.g., user identity code). Figure~\ref{fig:short_message_structure} shows the structure of a short message of 406 MHz beacon, while Figure~\ref{fig:long_message_structure} depicts the long message format. 
ELT's protocol code (37-bit to 40-bit) is 1000~\cite{cospas_specification}. 

The PLBs are designed for individuals such as SAR professionals, hikers, mountaineers, and seashore workers. This type of beacon uses both 121.5 MHz and 406 MHz, but COSPAS-SARSAT supports only 406 MHz. 
PLB's RF signal characteristics are as same as the ELT; however, the protocol code (37-bit to 40-bit) is different, for PLB the code being 1011~\cite{cospas_specification}.

The EPIRB is a maritime distress beacon device carried by vessels to alert SAR services to quickly locate the beacon/vessel in the event of an emergency, crash, or other distress situation. 
The COSPAS-SARSAT uses the 406.025 MHz channel for this service, while INMARSAT E listens at 1.6 GHz for EPIRB. The signal contains a 15-digit hex code of the beacon, country code, GPS position of the beacon, and others.
Differentiating from ELT and PLB, EPIRB's protocol code (37-bit to 40-bit) is 1010~\cite{cospas_specification}.

\section{Related Work}
\label{sec:relwork}

RF-based communication has always been a promising target for hackers because no physical tampering is needed. Additionally, the availability of required knowledge and technological improvement of hacking tools have aggravated security challenges. For example, similar to the COSPAS-SARSAT service, the aviation surveillance system Automatic Dependent Surveillance-Broadcast (ADS-B) was hacked already a decade ago~\cite{costin2012ghost,strohmeier2014realities}. The security vulnerabilities of a maritime surveillance service named Automatic Identification Systems (AIS) were also exposed a few years later~\cite{balduzzi2014security}.
Novel attacking concepts~\cite{khandker2021adsb,khandker2022ais,khandker2022security,turtiainen2022gdl90fuzz,juvonen2022apache} on these mission-critical surveillance systems (i.e., ADS-B, AIS, ACARS) have added a new layer of cybersecurity threats to both legacy and modern RF-based mission-critical infrastructure as well as deep beyond into the software stacks and potentially pivoting further into closely interconnected system elements. 
The mentioned studies show that using a Software Defined Radio (SDR), a targeted RF signal can be produced at a meager cost and effort~\cite{sruthi2013low,ferreira2020effective,dascal2013low,wright2019highly}. 
SDR also can \emph{easily manipulate} the encoded data of a radio signal (which otherwise is quite cumbersome for an embedded classical circuit), making the attack more rigorous and feasible. 

There have been several incidents of satellite hacking and exploitation. 
In Chaos Communication Camp 2015~\cite{iridium_hacking}, experts demonstrated how to hack the \emph{Iridium} network and eavesdrop on its pager traffic. 
\emph{SkyNet} satellites were reported to be hacked and asked for ransom~\cite{skynet_hacking}. 
Nowadays, state-sponsored satellite hacking is not a surprise. 
Russian hackers were blamed for hijacking commercial satellites to access western countries' sensitive diplomatic and military data~\cite{russian_hacker}. 
Following this, Ukrainian hackers claimed to have breached Russian \emph{Gonets-M} satellite systems~\cite{gonetsm}, as well as \emph{Satis} network's ground stations (running Yamal 401 and Ekspress-AM6 Satcom)~\cite{satis}. 
These reports assert that if there is any security loophole in the satellite communication system (SCS), considering the vast attack impact and financial and political gain, that could be a lucrative target for adversaries. Therefore, proper and updated security measures for an SCS are crucial. 
Yue et al.~\cite{Leosat_security} analyzed the security of LEO SCS. According to them, LEO SCSs are vulnerable to especially eavesdropping and malicious jamming, as these are relatively easy to carry out. LEO is gradually getting congested; approximately 4700 satellites are in this orbit. Therefore, inter-satellite interference is increasing. High-power radar, FM transmitters, aircraft, and others., from the ground also contribute to the interference. 
Yuqi et al.~\cite{cospas_interference} identified public walkie-talkies causing massive interference on COSPAS-SARSAT's uplink in China.
Pedersen et al.~\cite{Geosat_security} reported 15 threat sources for GEO SCS. They captured 400GB of data to analyze the real-life scenario. They found some data was not encrypted, which could be eavesdropped on effortlessly. Beside structural, environmental, and accidental issues, they believe professional hackers, national governments, competitors, and script kiddies could be interested in illicit activities on SCS.
Pavur~\cite{pavur-phd} studied cyber-physical security problems at the intersection of outer space and cyber-space for SCS. He divided satellite security into four sub-domain: radio communications security, ground systems security, space platform security, and mission operations security, and investigated all of them. He demonstrated that services from GEO satellites leaked sensitive data of many customers, including some of the world's largest corporations and critical infrastructure providers.

Several recent works touched on the cybersecurity aspects of the ``Emergency Services'' and ``Search And Rescue'' (SAR) systems. 
In~\cite{rieb2018case,lechner2019security}, the authors used ``Coordination Center East Thuringia: IT-Security in a Coordination Center'' as a case study, demonstrated the critical impact of cybersecurity risks and attacks for emergency services. 
Solcanu et al.~\cite{solcanu2021study} studied the effectiveness of using analog systems, such as amplitude (AM), frequency (FM), and phase (PM) modulations, in high-noise conditions in the marine environment for search and rescue missions. Their study aimed to evaluate whether the current emission classes are sufficiently resistant to disturbances or whether other technical solutions, such as the transition to digital communications, need to be adopted. They reported that the performance depends on the used emission classes; however, the AM system is generally more resilient to disturbance than FM or PM. 
Bernsmed et al.~\cite{bernsmed2021d4} introduced the concept of multi-modal communication in SAR operations in Norway. According to them, several means of communication technology such as AIS, VHF data exchange system (VDES), cellular, and satellite can be employed simultaneously or through a link of serial connections. All the stakeholders should be enrolled in the public key infrastructure (PKI) to ensure security. All actors can ignore the signature or disable encryption in emergency or poor communication conditions. When messages are relayed using different communication technologies, wrapping/tunneling needs to be used. 
Stavrinos et al.~\cite{stavrinos2022towards} studied the interoperability and cybersecurity of unmanned underwater vehicles in military/SAR operations. They identified that latency in underwater communication and the lack of standard communication protocols among different SAR parties are critical challenges for the internet of underwater things. They simulated three types of attacks. Firstly, a manual attack by eavesdropping on the communication using Wireshark. Secondly, using a semi-automated tool named Caldera, and finally, a fully automated attack using Infection Monkey. In all the cases, they demonstrated that the vulnerabilities lie in the existing system. The paper did not provide any practical solution; however, it suggested further investigating the vital issue of interoperability, standardization, and standard protocols.


\section{Methodology and Implementation}
\label{sec:meth-impl}

\subsection{Methodology}
\label{sec:meth}

As a basis, we take the very recently emerged and highly successful methodology developed and proposed by 
Khandker, Turtiainen, Costin, Hamalainen~\cite{khandker2021adsb,khandker2022ais,khandker2022security,turtiainen2022gdl90fuzz}. 
The authors successfully demonstrated the effectiveness of their approach on 
a wide range of systems, software, and devices related to aviation/avionics (ADS-B~\cite{khandker2021adsb,khandker2022security}, GDL90~\cite{turtiainen2022gdl90fuzz}) and maritime (AIS~\cite{khandker2022ais}) communications. 
Therefore, in line with the above, our methodology aimed at COSPAS-SARSAT EPIRB satellite communications can be similarly summarized as follows:
\begin{enumerate}

\item \textbf{Identifying} a critical communication protocol (i.e., COSPAS-SARSAT 406) that lacks minimal or robust cybersecurity features and protections (similar to ADS-B, AIS, ACARS)

\item \textbf{Coding/Decoding} (\textit{``codec''}) implementation of protocol specifications (i.e., COSPAS-SARSAT, EPIRB), for example, using Scapy~\cite{scapy-biondi2005packet} or ASN.1 tools

\item \textbf{Modulation/Demodulation} (\textit{``modem''}) and TX/RX implementation of protocol specifications (COSPAS-SARSAT EPIRB), compliant with used SDR hardware and software, for example, using GNU Radio~\cite{gnuradio-blossom2004gnu} or GNU Radio Companion (GRC)

\item \textbf{Cross-testing} that both \textit{``modem''} and \textit{``codec''} implementations work in all directions (within a safe and controlled environment, e.g., loopbacks within GNU Radio software itself, or unlicensed frequencies and Faraday-cage boxes/rooms)

\item \textbf{Generating input/output protocol packets} that correspond to the evaluated attack (e.g., replay, spoofing, DoS, ``coordinated attack''~\cite{khandker2021adsb,khandker2022ais})

\item \textbf{Subjecting the device/software/system under test} (e.g., EpirbPlotter~\cite{epirbplotter}, GEOLUT~\cite{geolut}, SARSAT, integrated software in SAR national centers) to the attack represented by the generated packets-stream

\item \textbf{Monitoring the device/software under test} for expected/unexpected results and any abnormal or non-compliant behavior

\item \textbf{Collecting and analyzing results}, to improve and fine-tune the attacks and the packets (e.g., packet contents, packet count), and then to restart from the \textbf{Generating input/output protocol packets} step

\end{enumerate}

\subsection{Implementation}
\label{sec:impl}

\paragraph*{Validating our Modulator and Coder}

Andy Walls presented an idea of transmitting the EPIRB signal~\cite{andywalls_grc_2016}. However, the author's work was limited to the simulation channel with random bit sequences. Inspired by his concept, we developed a GRC script to transmit valid SAR signals using transmission-capable SDR. 
We used Python programming language to create the bitstream of the SAR signal according to specified protocol~\cite{cospas_specification}. Bose–Chaudhuri–Hocquenghem codes (BCH codes) were calculated and inserted in the correct position. 
Later, the entire binary sequence was saved in a byte array. Our developed script takes this byte array as input and generates an RF signal as output. 
The developed software can encode any targeted information (e.g., country, location, beacon type) into a SAR signal. 

In a strictly controlled environment, we tested three transmission-capable SDRs (HackRF, BladeRF, and PlutoSDR) for RF signal transmission, and all supported the SAR signal transmission. Though one is enough for the test, we tested three SDRs to check the availability and support of devices in the ``attacking scenario''. 

We tested the reception of the transmission by a proprietary software called EPIRB plotter. Using RTL-SDR as the RF front-end, SDR sharp received the signal and generated the audio, which was subsequently fed into the EPIRB plotter using a virtual audio cable. The 406 MHz distress signals use a very low bit rate (400 bits per second). All the modern SDR receivers are capable of handling more than these bit-rates; hence the successful reception by cheap RTL-SDR devices indicates that the proposed \dumpFourZeroSix{} combined with RTL-SDR would be a feasible and affordable solution going forward.

\begin{figure}[!htb]
\centering
\includegraphics[width=0.90\columnwidth]{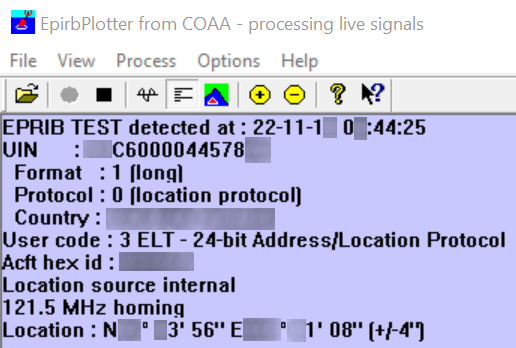}
\caption{Our \textit{spoofed} EPIRB-ELT signal (contains ICAO24 aircraft ID) well received by EpirbPlotter.}
\label{fig:elt}
\end{figure}

\begin{figure}[!htb]
\centering
\includegraphics[width=0.90\columnwidth]{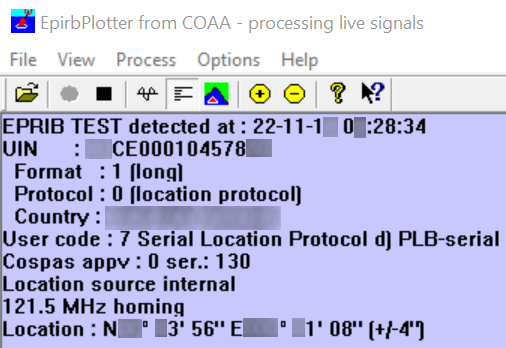}
\caption{Our \textit{spoofed} EPIRB-PLB signal well received by EpirbPlotter.}
\label{fig:plb}
\end{figure}

\begin{figure}[!htb]
\centering
\includegraphics[width=0.90\columnwidth]{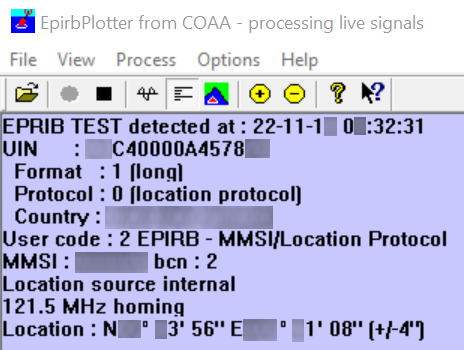}
\caption{Our \textit{spoofed} EPIRB-MMSI AIS signal (contains MMSI ship ID) well received by EpirbPlotter.}
\label{fig:epirb}
\end{figure}

\paragraph*{Validating our Demodulator and Decoder}

\begin{figure}[!htb]
\centering
\includegraphics[width=0.60\columnwidth]{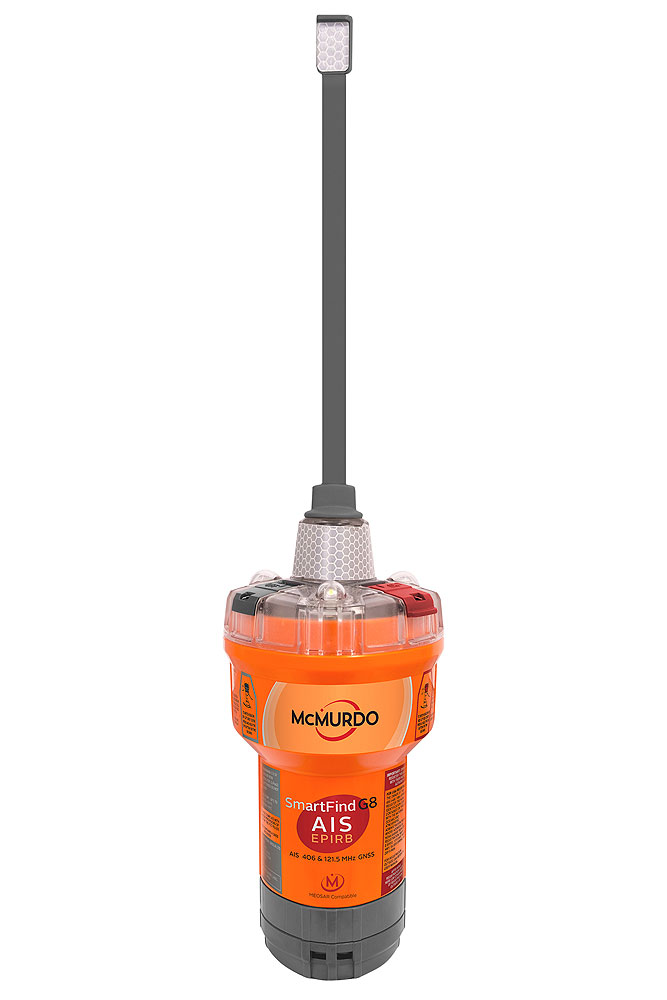}
\caption{Example of a compliant and certified ``test transmitter'' used in our labs and experiments -- a McMurdo G8~\cite{mcmurdo-g8}.}
\label{fig:mcmurdo}
\end{figure}

Besides our 406 MHz EPIRB transmit toolsets that we used for testing and demonstrating the attacks, we also developed 406 MHz EPIRB receive toolsets (i.e., demodulator and decoder) which we call \dumpFourZeroSix{}. 
We chose to implement the standard as a Scapy~\cite{scapy-biondi2005packet} class for the decoding part. 
We chose this approach as Scapy implementations can also be used effectively and efficiently in fuzzing protocols, thus potentially allowing us to find more cybersecurity issues in various devices and implementations. 
To test our 406 MHz EPIRB receive toolsets, we also used a McMurdo G8~\cite{mcmurdo-g8} (as in Figure~\ref{fig:mcmurdo}) in ``test mode''. 
McMurdo G8 device is handy for testing as it supports multiple beacons simultaneously -- 406 MHz COSPAS-SARSAT (EPIRB-MMSI) and 121 MHz AIS (AIS-MMSI) -- all underpinned but high-accuracy GNSS location information. 

\section{Attackers, Threats, Exploits, Challenges}
\label{sec:attacks}

\subsection{Attacker Model}
\label{sec:attackers}

The overall threat model for COSPAS-SARSAT could be seen as a similar and generalized form of the attacker model by Costin et al.~\cite{costin2012ghost}, and could be summarized as:
\begin{enumerate}
\item \textbf{Under-attack system} elements (such as global COSPAS-SARSAT EPIRB elements -- satellites, ground stations, user terminals, software running on those elements,  and others.) are authentic, authorized and benign (i.e., not hosted or owned for malicious purposes), hosted on an original, trusted, or hardened infrastructure or devices (i.e., a trusted computing base, including OS, UI, web server, interpreter, and others.), and operated by non-malicious stakeholders
\item \textbf{Attacker} has access to (and understanding of) the protocol specifications, e.g., COSPAS-SARSAT~\cite{cospas-406-distress-specs}
\item \textbf{Attacker} has minimal SDR and software programming skills
%
\item \textbf{Attacker} has a minimal budget (e.g., less than 1000 EUR/USD) to acquire basic yet powerful and flexible SDR tools (e.g., HackRF, BladeRF) and affordable RF power amplifiers (e.g., Mitsubishi RA07M4047M)
\item \textbf{Attacker} can emit the minimal power required according to specification, e.g., from 25 mW (sweeping-tone signal constantly on 121.5 MHz) up to 5 W (burst about every 52 seconds at 406 MHz)
\item \textbf{Attacker} can freely manipulate either the contents or the sequence of COSPAS-SARSAT EPIRB communication packets requests and responses sent to the original COSPAS-SARSAT receivers/terminals, but cannot or did not directly compromise the infrastructure or the application/firmware code prior (namely, supply chain attacks)
\end{enumerate}

\subsection{Threats Model}
\label{sec:threats}

Below we shortly enumerate the possible threats enabled by the fundamental lack of cybersecurity controls built into the 406 MHz COSPAS-SARSAT protocols (e.g., EPIRB). 

\begin{enumerate}

\item \textbf{Basic Replay.} The attacker captures raw COSPAS-SARSAT signals (e.g., I/Q) on 406 MHz using affordable and highly-available devices (e.g., RTL-SDR, HackRF, FlipperZero) and subsequently sends/replays the raw signals to RF. The attacker does not need to code any software, as replaying does not require coding/decoding of packets nor modulation/demodulation of signals. 
\textbf{We have confirmed this attack on COSPAS-SARSAT in the lab.}

\item \textbf{Basic Spoofing.} The attacker can spoof (either partially or fully) any valid and legitimately looking COSPAS-SARSAT messages. The receiver cannot detect that such spoofing is occurring without having another means of verification (e.g., contacting the sender via phone or the internet). Two-way verification is problematic as the COSPAS-SARSAT is a distress/SAR technology and likely alternative communication is unavailable with the sender). 
\textbf{We have confirmed this attack on COSPAS-SARSAT in the lab.} (Section~\ref{sec:impl})

\item \textbf{Close-to-target Mimicry Spoofing.} The attacker can deploy a spoofing device near the target, e.g., using a cheap drone nearby the target or even physically attached to the target. 
For example, the target could be a private jet (EPIRB-ELT attack vector -- Figure~\ref{fig:elt}), a private yacht (EPIRB-MMSI attack vector -- Figure~\ref{fig:epirb}), or firefighting or sea-rescue machinery (EPIRB-MMSI attack vector -- Figure~\ref{fig:epirb}) and personnel (EPIRB-PLB attack vector -- Figure~\ref{fig:plb}). 
Due to the ``close-to-target'' setup, upon a COSPAS-SARSAT spoofed signal being issued, it is hard/impossible for the SAR Coordination Center to know that the signal is indeed spoofed. 
From the point of view of the SAR Coordination Center treating the alert, they know the target is supposed to be precisely where the spoofed signal is sent from and could/should be treated as an actual distress signal, thus triggering the whole SAR operation chain of events, commands, and actions. 
This attack is even harder to detect under harsh weather and low-visibility conditions, e.g., large wildfire or storms, where it is impossible to confirm the situation visually. At the same time, the signal cannot be ignored due to ``spoofing suspicion'' -- what if, indeed, a firefighter or a sea rescuer in distress needs help in a zero-visibility area?

\item \textbf{Overwhelming Spoofing.} The attacker overwhelms the entire COSPAS-SARSAT system with spoofed signals, which can occur globally, nationally, or regionally -- depending on the attacker's goals, motivations, capabilities, and resources. 
For example, this can be accomplished realistically with an army of cheap drones carrying a COSPAS-SARSAT spoof device at the exact GPS locations where the signal is desired to be spoofed. Geolocation accuracy is required, as the satellites use multi-lateration (MLAT) on the source signal, and any discrepancy with the GPS location encoded in the packets could be easily used to detect and flag these as spoofed signals. 
Though it may be apparent to the SAR Coordination Center that a cyberattack is ongoing, it may be hard to distinguish real vs. spoofed signals given the overwhelming number of SAR signals incoming, thus posing a risk to SAR resource exhaustion or wrongly prioritizing the SAR targets and missions. 

\item \textbf{``Overwhelming + Close-to-target Mimicry'' Spoofing.} This attack is a combination of the two attacks above. 
From the attacker's perspective, it will be most effective and devastating when a massive incident is ongoing (e.g., massive wildfires or sea rescue operations). The attacker would spoof the COSPAS-SARSAT signals (e.g., EPIRB) with correct IDs associated with each rescuer and machinery involved, essentially triggering a recursively-amplified denial-of-service (DoS) attack on the SAR system and operational architecture itself. 

\item \textbf{Network/Application Fuzzing and Exploitation.} The attacker uses well-known and traditional fuzzing and penetration testing techniques (at the network and/or application layers) to find additional design and implementation vulnerabilities in the COSPAS-SARSAT software stacks. 
In this sense, the most promising results and techniques related to aerospace and maritime were recently developed and demonstrated in~\cite{khandker2021adsb,khandker2022ais,khandker2022security,turtiainen2022gdl90fuzz,juvonen2022apache}. 
We successfully replicated this technique to COSPAS-SARSAT, but due to the deficient number of COSPAS-SARSAT software accessible to us (i.e., only EpirbPlotter), we are unable to have a strongly conclusive result for the entire COSPAS-SARSAT ecosystem, except the fact that we were unable to crash EpirbPlotter at this point. 
However, our COSPAS-SARSAT over-the-air and protocol fuzzing techniques could be instrumental in discovering critical vulnerabilities and crashes in official COSPAS-SARSAT software used in SAR Coordination Centers. 
\textbf{We have implemented this attack, and our immediate future work aims to focus on expanding this direction.}

\end{enumerate}

It is a matter of time before some or all of the above threats could be employed in active cyberwarfare as a part of Electronic Warfare (EW) and tactical resource drain/lockup from other strategic missions. For example, when SAR missions are activated based on the enemy's fake COSPAS-SARSAT legitimately-looking spoofed signals.

\subsection{Weaknesses and Mitigations}
\label{sec:weakness}

Below we shortly enumerate the primary and most essential weaknesses of the current technical specification and operation of the entire COSPAS-SARSAT system. 

\begin{enumerate}

\item \textbf{Need for Message Authenticity.} Lack of secure digital signatures for message and protocol authenticity to prevent spoofing attacks.

\item \textbf{Need for Message Freshness.} Lack of random unique non-reusable ``nonce'' sequences/tokens in messages or protocols that are not based on ``challenge-response'' to prevent replay attacks.

\item \textbf{Need for Randomized and Confidential IDs.} Generate the IDs of each COSPAS-SARSAT device (e.g., EPIRB, PLB, ELT) in a random manner (i.e., non-sequential, non-predictable). Moreover, keep these IDs confidential (as part of data protection and sensitivity planning), so potential attackers cannot easily find them, thus making their spoofing attempts slightly less successful. 

\end{enumerate}

The most likely (and reasonable) explanation for the above is that at the design and implementation stages (i.e.,  in the1980s--1990s) of the COSPAS-SARSAT systems, the attacker model was considerably different and assumed that it would require state-level attackers to be able to communicate and interfere with space and satellite systems.
However, the rapid technological developments in the 2000s and the ``explosion'' of affordable yet powerful SDR toolsets changed the attacker model radically; thus, the COSPAS-SARSAT system as-is today became vulnerable. 
Similarly to vulnerable ADS-B, AIS, and ACARS systems, replacing the protocol in such a global and evolved system is highly unlikely due to multiple challenges (e.g., complex legacy architecture, costs, and budgets, impact on the availability of the system, integration of legacy and new parts, deployed satellites hard to replace). 
However, possible mitigations and workarounds could be similar to implementing encryption or digital signatures on top of vulnerable ADS-B~\cite{Yang2019ADSB,Wu2020adsb_signature}, AIS~\cite{Goudossis2019AIS,Sciancalepore2021AIS} deployments. 
We leave the exploration and practical research of similar defensive solutions on top of existing COSPAS-SARSAT protocols and implementations as immediate future work.

%

\subsection{Feasibility of weaponizing drones and orbiting objects}
\label{sec:weaponize}

Recently, Abedi et al.~\cite{abedi2020wifi,abedi2022non} demonstrated practical and feasible implementation of small cheap drones carrying hardware capable of exploiting Wi-Fi security and privacy vulnerabilities. 
On a similar concept line, any such drone can be repurposed or enhanced to carry hardware and software for cyber-exploitation (e.g., DoS, RCE) or spoofed signals, including COSPAS-SARSAT (Section~\ref{sec:threats}) as well as ADS-B~\cite{khandker2021adsb,khandker2022security}, and AIS~\cite{khandker2022ais}. 
Moreover, orbiting objects (e.g., satellites, crew ships, rockets) of both nation-states and commercial organizations could be weaponized similarly.
In the end, such weaponized drones and orbit objects could soon (or already!) carry cybersecurity payloads for a wide range of space/satellite systems, including COSPAS-SARSAT, thus targeting to exploit either the operational functions (Section~\ref{sec:threats}), or the network and software functions~\cite{costin2012ghost,turtiainen2022gdl90fuzz,khandker2021adsb,khandker2022ais,khandker2022security,juvonen2022apache}. 
For example, a recent article features a Space Force general detailing how jamming, blinding lasers, cyber-attacks, and other satellites have America's space-based capabilities under siege~\cite{space-force-general}.

\subsection{Core Research Challenges}
\label{sec:chal}

During our experiments, we identified core challenges limiting or preventing certain types of experiments or validations related to cybersecurity weaknesses and potential exploitation of the COSPAS-SARSAT protocols and systems. 
We will systematize the main challenges below.

\begin{enumerate}

\item \textbf{Highly-sensitive technology.} Indeed, this technology is sensitive, as any erroneous use of the technology or any tests that go wrong may impact the real world and use highly-expensive and critically-scarce resources for no good reason. 
This makes research-based experimentation with the technology a high-risk activity. 

\item \textbf{Limited access to systems and software.} Indeed, acquiring software and hardware that support COSPAS-SARSAT protocols is either cost-prohibitive or is strictly controlled by the seller, where the systems are sold only to government-licensed SAR centers and organizations. 
This makes research-based experimentation with the technology highly limited. 

\item \textbf{Limited access to (and engagement with) stakeholders.} Indeed, the government-licensed SAR centers and organizations have a clear mandate, and their power of decision regarding freestyle experimentation with SAR is limited and comes with significant liability. 
This makes research-based experimentation with the technology highly limited. 

\end{enumerate}

As COSPAS-SARSAT is a space-/satellite-based technology, and at the same time, it is a critically important search and rescue global technology, access to these devices and software is minimal and well-controlled. Thus even ethical and strictly controlled experimentation is not risk-free or readily available. 
As immediate future work, we aim to establish national and international contact points with COSPAS-SARSAT centers to bootstrap cybersecurity readiness testing and exercises involving some of the attacks we presented above. 
Moreover, we cordially invite any COSPAS-SARSAT national centers to engage with us to establish cybersecurity collaborations, practices, and periodic assessments of such critical space technology. 


\section{Discussion and Future Work}
\label{sec:discuss}

\subsection{406 MHz open-source software}
\label{sec:opensource}

As detailed in Section~\ref{sec:impl}, one side-effect of our experiments is that we developed and proposed \dumpFourZeroSix{} (an equivalent to famous ADS-B-focused \dumpTenNinety{}) aimed at COSPAS-SARSAT 406 MHz transmissions. For example, we believe that \dumpFourZeroSix{} can be effectively and efficiently used for research, amateur, and crowdsourcing projects and to support global augmentation, collection, and analysis of COSPAS-SARSAT and related aerospace data for both cybersecurity and data analytics applications. 
We plan to release \dumpFourZeroSix{} receiving toolset as open-source in order to support further research and experimentation. 

Recently, Mladlenov et al.~\cite{mladenov2022implementation} demonstrated the GNU radio-based SAR receiver's implementation and porting it to an operational environment for onboard deployment on OPS-SAT. However, they did not release source code, hindering the open-source and crowdsourcing efforts.
Moreover, the fact that their implementation aims to run on a satellite is quite a restrictive factor to the entire research and enthusiasts community (except the select few with access to deployed satellites and lab flatsats). 
Finally, the main difference between~\cite{mladenov2022implementation} and our present work is that~\cite{mladenov2022implementation} is limited to \textit{receive-only} mode while aiming at functional exploration of COSPAS-SARSAT. In contrast, our present work covers reception and transmission while aiming specifically at cybersecurity explorations and aspects.

\subsection{Future research, open-data and crowdsourcing}
\label{sec:opendata}

To our best knowledge, no free nor open-source software is available to decode SAR signals. Therefore, one of our research goals was to develop a freely available \dumpFourZeroSix{} software similar to the famous \dumpTenNinety{} for ADS-B. 
The development and distribution of \dumpFourZeroSix{} would help receive the SAR signals in a crowdsourced manner. 
One good example candidate for such enhancement and integration is the famous OpenSky Network~\cite{strohmeier2015opensky}. 
As a result, the data can also be collected via crowed-sourced nodes, thus not relying on the open-data policies of the COSPAS-SARSAT centers and operators. 
In this way, the \dumpFourZeroSix{} would broaden the reception platforms (whether COSPAS-SARSAT or OpenSky), and the substantial amount of geographically-specific crowdsourced data could be used to build a statistical model to identify the probability of both real and spoofed SAR alerts, timing, and prediction. 
The statistical model may also be used to compare the received signal strength against the claimed position, thus detecting possible spoofing.

As possible future work, we plan to experiment with and research the (strong) cryptographic support for lightweight COSPAS-SARSAT implementations. For example, we plan to investigate the feasibility of using digital signatures and lightweight (public-key) crypto, whether as an extension to the standard or as a specification bypass (e.g., using future/unused fields).


\section{Conclusion}
\label{sec:concl}

In this paper, we investigate the cybersecurity aspects of COSPAS-SARSAT space-/satellite-based systems. 
In particular, we practically and successfully implement and demonstrate the first (to our knowledge) attacks on COSPAS-SARSAT 406 MHz protocols: replay, spoofing, and protocol fuzzing. 
We also identify a set of core research challenges preventing more effective cybersecurity research in the field and outline the main cybersecurity weaknesses and possible mitigations to increase the system's cybersecurity level. 
Moreover, we developed and proposed \dumpFourZeroSix{} (an equivalent to famous ADS-B-focused \dumpTenNinety{}) aimed at COSPAS-SARSAT 406 MHz transmissions. For example, we believe that \dumpFourZeroSix{} can be effectively and efficiently used for research, amateur, and crowdsourcing projects and to support global augmentation, collection, and analysis of COSPAS-SARSAT and related aerospace data for both cybersecurity and data analytics applications. 


\section*{Acknowledgments}
\label{sec:ack}

Minor sections and some hardware of this research were kindly supported by the cascade funding from 
Engage KTN (SESAR Joint Undertaking under the European Union's Horizon 2020 research and innovation programme under grant agreement No 783287) 
project \emph{"Engage - 204 - Proof-of-concept: practical, flexible, affordable pentesting platform for ATM/avionics cybersecurity"}. 
All and any results, views, opinions are authors' only and do not reflect the official position of the European Union (and it's organizations and projects, including Horizon 2020 program and Engage KTN). 
Major parts of this research were supported by 
\emph{``Decision of the Research Dean on research funding (20.04.2022)''} 
within the Faculty of Information Technology of University of Jyv\"{a}skyl\"{a}. 
Hannu Turtiainen also thanks the Finnish Cultural Foundation / Suomen Kulttuurirahasto (https://skr.fi/en) 
for supporting his Ph.D. dissertation work and research (under grant decision no. 00221059) and the Faculty of Information Technology of the University of Jyv\"{a}skyl\"{a} (JYU), 
in particular, Prof. Timo H\"{a}m\"{a}l\"{a}inen, for partly supporting and supervising his Ph.D. work at JYU in 2021--2023.
Syed Khandker was partially supported by the Finnish Foundation for Technology Promotion under the PoDoCo grant program. 
The authors also thank Ahsan Saleem (University of Jyv\"{a}skyl\"{a}) and Laura Tirri (University of Jyv\"{a}skyl\"{a}) for their contributions. 



\tiny{
\bibliographystyle{IEEEtran}
\bibliography{IEEEexample.bib}
}

\end{document}